# Modeling of Branched Thickening Polymers under Poiseuille Flow Gives Clues as to how to Increase a Solvent's Viscosity

E. Mayoral[1] and A. Gama Goicochea[2*]

[1]Instituto Nacional de Investigaciones Nucleares, Carretera México Toluca s/n, La Marquesa Ocoyoacac 52750, Estado de México, Mexico

[2]Departamento de Ingeniería Química y Bioquímica, Tecnológico de Estudios Superiores de Ecatepec, Ecatepec de Morelos 55210, Estado de México, Mexico

## ABSTRACT

The viscosity enhancement of a solvent produced by the addition of thickening branched polymers is predicted as a function of polymer concentration, branch length and persistence length, and strength of the covalent bonding interactions. Non equilibrium, stationary state Poiseuille numerical simulations are performed using the dissipative particle dynamics model to obtain the viscosity of the fluid. It is found that the clustering of the polymers into aggregates raises the viscosity and that it is more strongly affected by the strength of the bonding interactions. General scaling relationships are found for the viscosity as a function of the variables studied, which are expected to be useful for the design and synthesis of new viscosifying polymers. It is argued that our results can be applied to aqueous thickeners, of importance for colloidal fluids such as paints and coatings, and also for nonpolar fluids such as supercritical $CO_2$, which is a promising non – hydraulic fracking fluid also useful in enhanced oil recovery.

---

[*] Corresponding author. Electronic mail: agama@alumni.stanford.edu

# INTRODUCTION

Research related to the design of new thickening additives that can increase the viscosity of fluids efficiently at low concentration, that are environmentally friendly and inexpensive is a topic considerable current interest. These additives are used, for example, to increase the viscosity of different solvents such as water, to fine – tune the rheology of water – based paints [1], or for supercritical $CO_2$, which can be used as a water substitute in fracking and for enhanced oil recovery [2 – 4]. Thickening agents are typically polymers added to a fluid to raise its viscosity without changing the rest of its properties significantly. Thickeners are also used to treat industrial water to make it reusable in processes and as clarifiers to remove fine minerals from the water [5, 6]. However, one of the areas where the design and synthesis of new viscosifiers has gained the greatest attention recently has been in the petrochemical sector, specifically in enhanced oil recovery (EOR) and in fracking [2 – 4, 7, 8]. Nowadays EOR processes have become especially important to the overall oil production. Some fluid – based techniques belong to these processes. Among these methods, $CO_2$ flooding has been tested successfully, showing to be effective not only in EOR but also as an agent that reduces greenhouse emissions [9, 10]. Supercritical $CO_2$ is particularly promising as a fracturing fluid because of its liquid – like density, however its viscosity is gas – like, which can be two orders of magnitude lower than the viscosity of water [11]. Low viscosity leads to viscous fingering and low oil mobility [2, 10], limiting the applications of $CO_2$ as a fracturing fluid. One strategy to thicken $CO_2$ that has received considerable attention recently is the addition of polymers to it; using this method the viscosity of $CO_2$ can be increased by more than two orders of magnitude using fluorinated copolymers [5]. Consequently, the capability to regulate and enhance the viscosity of $CO_2$ flooding fluids has impact on energy production.

In complex fluids, colloidal aggregates can improve the viscosity of a solvent, forming interconnections or associations in the bulk phase [12]. Researchers have tried to increase the viscosity of $CO_2$ adding polymers that are able to form intermolecular associations in solution [13]. Common compounds known to thicken oils such as hydrocarbon – based polymers, telechelic ionomers, organometallic compounds, surfactants, and ammonium carbamates have extremely low solubility in $CO_2$ and water [13]. In these solvents, viscosity improvement using traditional thickeners has been achieved when an organic cosolvent like ethanol or toluene is added [14] to improve their solubility. The first high molecular weight polymers capable of dissolving in $CO_2$ at moderate pressures without the need for a cosolvent were reported by DeSimone and coworkers [15]. Huang and co-workers [5] enhanced the viscosity of $CO_2$ by up to 400% using styrene – fluoroacrylate copolymers at a relatively large concentration of 5 wt %. These copolymers contain a hydrocarbon backbone, having dangling branches of $C_8F_{17}$ and shorter branches of styrene [5]. The function of styrene is to promote intermolecular interactions between the aromatic rings in one copolymer chain and those in another (π – stacking interactions). Fluorine increases the solubility in $CO_2$, which is achieved because functional groups such as fluorocarbon chains interact favorably with the quadrupole moment of $CO_2$ [16]. However, the environmental impact and high cost of this kind of perfluoro polymers prevent their large-scale use [10, 17]. On the other hand, it has been shown that highly branched hydrocarbons, specifically those containing *t*-butyl (–$C(CH_3)_3$) [18, 19] are also compatible with liquid $CO_2$. Additional structures which can improve the solubility in $CO_2$ by complementary Lewis – base interactions are C=O and –C–O–C– oxygenated components [20 – 23].

Extensive research has been carried out to control the thickener solubility in different solvents, modifying the thickening polymers' chemical structure, introducing different functional groups to promote a solvent – soluble tail group association [2]. The next challenge is to enhance the performance of the thickener, modifying its architecture or using co – additives and cosolvents to make formulations friendlier with the environment. To substantially increase the viscosity of a solvent when thickening agents are present in dilute concentration, a promising option is creating macromolecular supra structures via noncovalent associations. It has been reported [24, 25] that random, lightly sulfonated polystyrene is able to significantly improve the solution's viscosity in nonpolar organic solvents via the union of the acid or salt groups in the polymeric chain. In particular, this mechanism is displayed by ionic polymers, known as ionomers [26], which are polymers containing minor quantities (up to 15 mol %) of ionic groups. Compared with nonionic polymers, these ionomers at low concentration display uncommonly elevated thickening performance in nonpolar solvents [24, 27]. The addition of small amounts of polar cosolvent to improve solvation of ionic groups in these systems reduces the viscosity of the solution [27]. The viscosity is also reduced in experiments where the temperature is changed, creating a different viscosity – temperature relationship from those exhibited by typical polymer solutions [26, 27]. This anomalous behavior is attributed to the fact that the ionic groups are essentially non – ionized in solvents of low dielectric constant, so the solution performance is dictated by ion pair interactions rather than by free ions.

Although most works find that the rheological properties of fluids with thickening polymers are defined by the intertwining between polymer chains [28], the detailed mechanisms that produce those associations are not sufficiently well known. Molecular simulation [29 – 33] plays a key role in understanding the association mechanisms of polymers in fluids, because different polymer architectures can be modeled, as well as bonding interactions and concentrations. The flexibility of the polymer chain is another important variable to control in the design of new thickeners, which can be modeled also in molecular simulations. It is known that at higher temperatures the energy barrier of the C–C bond rotation can be overcome, allowing the possibility of steric rearrangements to establish interconnected structures [2]. Atomistically detailed molecular simulations [29 – 33] are very accurate, but still require long time to solve the cooperative motion of many polymer chains immersed in a fluid. To reduce simulation times and reach larger length scales, coarse – grained, mesoscale methods are a useful alternative. One of the most commonly used mesoscale simulation techniques is dissipative particle dynamics (DPD) [34 – 36], whose simple force fields and stable thermostat make it ideal to study many polymers in solution. The forces that make up the DPD model are pairwise additive, therefore the local and global momenta are conserved. This feature ensures that all the hydrodynamic modes are preserved [35]. In a recent work [37], linear and ring polymers under pressure – driven flow were modeled also. However, the hydrodynamic interactions were artificially added to the dynamics of the fluid by a so – called multiparticle collision algorithm (MPC). It is claimed that the MPC conserves "local energy, momentum and mass" [37]. In the DPD model this is unnecessary since all its forces are pairwise additive, therefore they also conserve the local and global momenta and obey the Navier – Stokes equations [35].

In this work we use DPD to predict the viscosity of branched polymers in a solvent under stationary Poiseuille flow. The variables studied are the length of the polymers' branches, the strength of the bonds between monomers, the intensity of the angular interaction between neighboring bonds (which determine the flexibility of the polymers), and the polymer concentration. Our aim is to provide clues for the design and synthesis of optimal thickening molecules that can increase the viscosity of fluids such as $CO_2$ for non-hydraulic fracturing [5], and aqueous complex fluids, such as paints [1]. This article is divided into four sections. After this Introduction, the DPD model, the simulation protocol and its details are presented in the section Models, Methods and Simulation Details. The structure of the modeled polymers and their bonding and non-bonding interactions are described in that section as well. In the Results and Discussion section we show the effectiveness of the thickeners tested, calculating the viscosity of the fluid under Poiseuille flow. This type of flow, where an external force acts on all the molecules of the fluid, is chosen because it is the one that most closely resembles the type flow exerted in applications of technological interest such as in EOR. We summarize our findings in the Conclusions section.

## MODELS, METHODS AND SIMULATION DETAILS

Numerical simulations using DPD model [34 – 36] are performed for a model fluid with thickening model polymers under stationary Poiseuille flow. DPD is a well – known, particle – based approach where coarse graining is applied to solve the motion of the particles in the system whereby significant gains in computational time and system size are achieved. DPD simulations have been successfully applied to various systems [38] and have been capable to predict qualitatively and quantitatively the equilibrium and non – equilibrium properties of electrostatic and neutral complex fluids [39]. Since DPD is relatively well known now, only the essential details that are relevant for the present work shall be presented here. More details about the fundamentals and applications of the DPD can be found in recent reviews [40,41].

In DPD simulations, the atoms or molecules in the system contained in a volume element are grouped into spherical beads of the same size, forming clusters of complex fluid components for which the equation of motion is solved. The coarse-grained system is made up of a set of $N$ particles or DPD beads; each one is characterized by its position $\vec{r_i}$ and momentum $\vec{p_i}$. Solving Newton's second law at discreet time steps yields the evolution of positions and momenta of all particles as time evolves [42]. The DPD interparticle force applied by the particle $i$ on the particle $j$ is pairwise additive and has three components: conservative $\vec{F_{ij}^C}$, dissipative $\vec{F_{ij}^D}$ and random $\vec{F_{ij}^R}$. The effective force on the $i$-th particle, $\vec{F_i}$ is given by:

$$\vec{F_i} = \sum_{i \neq j}[\vec{F_{ij}^C} + \vec{F_{ij}^D} + \vec{F_{ij}^R}]. \tag{1}$$

All force short ranged; the amplitudes of the dissipative and random forces are not independent, they are linked through the fluctuation–dissipation theorem, which produces a built-in thermostat [35]. The conservative force is given by the following expression:

$$\vec{F_{ij}^C} = a_{ij}\left(1 - \frac{r}{r_c}\right)\widehat{r_{ij}} \tag{2}$$

where $\widehat{r_{ij}}$ is the unit vector considering particles $i$ and $j$, $r$ is the relative position between them, $r_c$ is the cutoff distance and $a_{ij}$ is the strength of the repulsive force between the beads. The parameter $a_{ij}$ is obtained from the Flory-Huggins parameter [36]. Polymer chains are formed using the Kremer – Grest bead – spring model [43], where a harmonic force joins the neighboring beads along the chain and is given by:

$$\vec{F_S}(r_{ij}) = -k(r_{ij} - r_o)\widehat{r_{ij}}, \tag{3}$$

with $r_o$ being the equilibrium distance of the spring and $k$ is the spring constant. Usually the value for $r_o = 1/\rho^{1/3}$ is used so that it is independent of the stiffness of the polymer spring [44, 45]. In our simulations the number density $\rho$ is set equal to 3, then $r_o = 0.7$. Various values for the spring constant have been used in the literature [46 – 48]. However, the common practice is to choose its value such that the equilibrium spring distance $r_o$ corresponds to the distance where the first maximum in the pair correlation function of non – bonded beads is found [46, 47, 49]. The effect of systematic variations in the parameters that characterize the spring force between bonded beads on interfacial tension and excess pressure has been studied by one us [49]. The results showed that $r_o = 0.7$ and $k = 100$ are optimum values that guarantee that the excess pressure and the interfacial tension are insensitive to changes in these parameters [49]. Additionally, a recent study shows that it is possible to characterize incompletely solvated ionic surfactants with a weak spring constant [50]. These works suggest that values of the spring constant $k$ can be chosen to mimic weak or strong particle bonding. This is one way of introducing into the model the possibility of including different intensities for intramolecular noncovalent and non – electrostatic secondary bonds or intermixing.

Since the purpose of this work is to test the viscosifying capabilities of polymers with common general characteristics so that our findings can be of use for various polymer/solvent environments, no attempt was made to tailor our simulations to a specific chemical structure. However, there are reports showing that branched polymers are better thickeners than linear ones because branching increases their solubility and promotes intermolecular association [2, 5]. For this purpose, we model prototype star – like polymers with four arms (BR$_4$), where the central structure is made up of six beads labelled as B. Structures similar to this can be found in the literature [51, 52]. The four arms are made up of beads called R; see Fig. 1(a). The mesoscopic DPD representation of the system simulated is shown in Fig. 1(b), while Table 1 shows the $a_{ij}$ interaction parameters, see Eq. (2). They were chosen using the standard method [36] for a coarse – graining degree equal to three, i.e., a DPD bead has the volume of three water molecules.

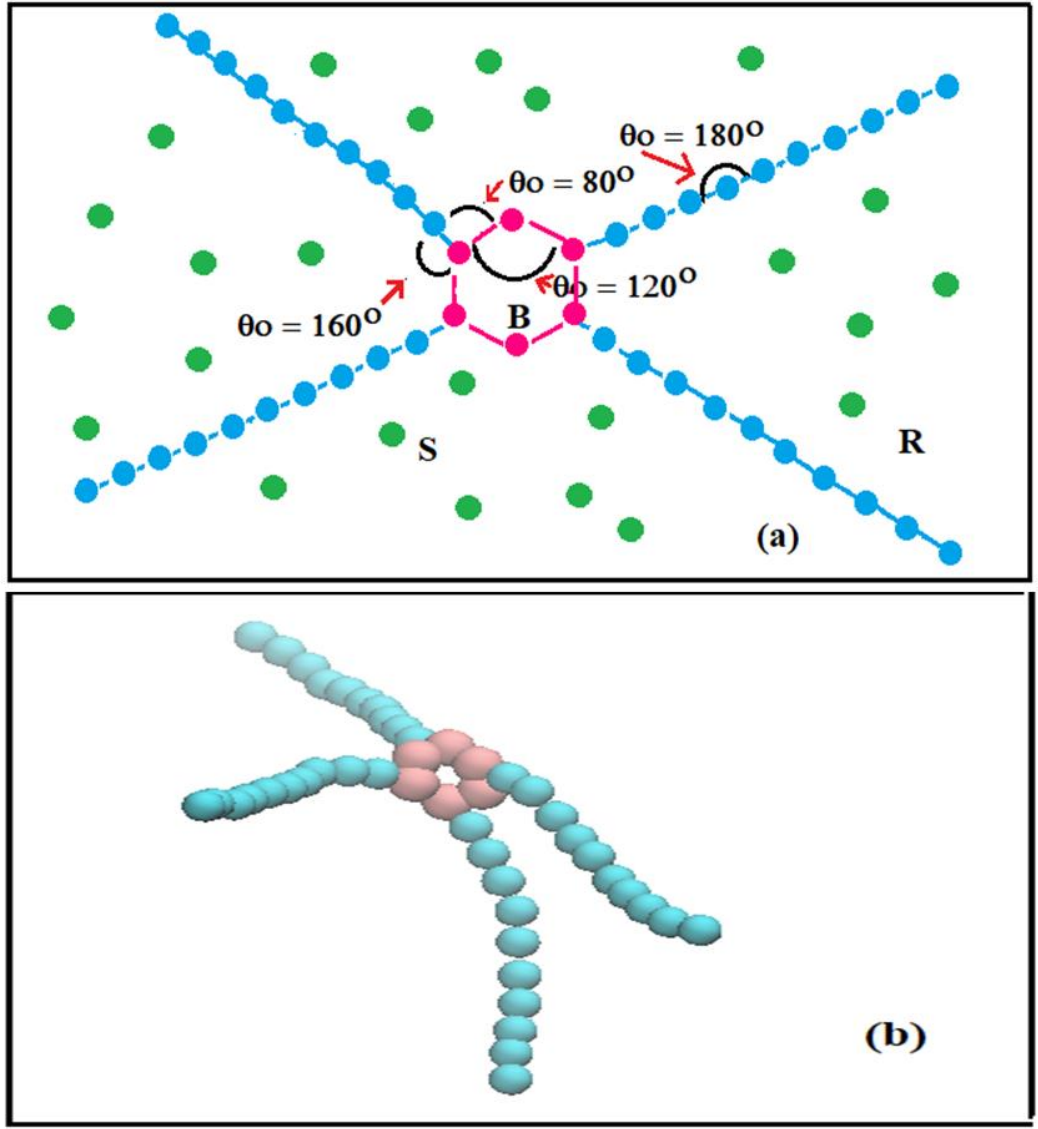

**Figure 1**. (a) Schematic representation of the structure of the DPD prototype copolymer BR$_4$ (thickener) in a solvent (S); $\theta$ is the angle between bonds, which determines the stiffness of the branches, see Eq. (4). R represents the branches and B the central structure. (b) Snapshot of a copolymer extracted from our DPD simulations. Beads in cyan make up the branches and those in pink represent the central structure of the copolymer. For the case shown, the branches have polymerization degree equal to $N = 10$. The solvent is not shown for simplicity.

The flexibility of the arms is modeled by a harmonic bond angle interaction:

$$F_A = -k_\theta(\theta - \theta_0) \tag{4}$$

where $k_\theta$ is the strength of the angular spring, $\theta$ is the angle between bonds and $\theta_0$ is the equilibrium angle between the bonds, see Fig. 1(a).

**Table 1**. Conservative DPD interaction parameters $a_{ij}$ used in most of this work, except where stated otherwise, see Eq. (2). S represents the solvent beads; B beads make up the central structure of the BR$_4$ polymer and R beads are those forming the branches. These parameters are chosen so that theta solvent conditions are obtained for the branches. See also Fig. 1.

| $a_{ij}$ | S    | B   | R    |
|----------|------|-----|------|
| S        | 78.3 | 100 | 79.3 |
| B        |      | 50  | 100  |
| R        |      |     | 79.3 |

The parameters listed in Table 1 were chosen so that good solvent conditions between the beads along the polymer branches, R, and the solvent beads, S, were achieved, while theta conditions are chosen for B beads with solvent beads. This promotes the solubility of the branched polymer in the solvent, to map what is found in experiments and simulations [5, 10, 53]. The B – B interaction between beads forming the central structure of the star – like polymers was chosen to mimic π – π stackings, as reported for thickening polymers with aromatic rings that improve intermolecular interactions, which in turn yield larger viscosity [5]. All simulations were carried out under the canonical ensemble, i.e. at constant number density and temperature. The total number of beads in all cases was $1.5 \times 10^4$; the side of the cubic simulation box is $L = 17.0 r_C$ so that the total number density is $\rho = 3/r_C^3$. All simulations were run for at least 15 blocks of $10^4$ time steps each, with the first seven blocks used to reach stationary flow, and the rest for the production phase. Reduced units are used throughout this work, with the reduced time step $\delta t^* = 0.03$ and $m^* = T^* = r_C^* = 1$ for the reduced mass, temperature and cutoff radius, respectively.

To calculate the shear viscosity, simulations were run under stationary, Poiseuille flow [54] using the setup shown in Fig. 2(a), where two parallel walls confine the fluid perpendicularly to the $z$ – axis. This type of flow is obtained in this work by applying an external constant force on all particles in the fluid along the $x$ – direction only. The two fixed walls with total separation distance $L = 2d = 17r_C$ are modeled as short – ranged, linearly decaying forces acting along the $z$ – axis given by the force law $F_w(z) = a_w[1 - z/z_C]$, for $z \leq z_C = 0.4r_C$, and zero otherwise [55]. Here, $a_w$ is the wall force amplitude, $z_C$ is the maximum range of the wall force, and $r_C$ is the DPD cutoff length. For all the simulations reported in this work the value $a_w = 115.0$ (in reduced units) was used, except where stated otherwise. The use of this setup removes the need to resort to higher density walls of frozen particles because the DPD beads do not penetrate the walls [56, 57]. Also, artifacts introduced in the velocity profiles when using the Lees−Edwards periodic boundary conditions [58, 59] are avoided. The featureless walls used here have the extra improvement of having a controllable range, through the $z_C$ parameter, helping to reduce the slip length [60]. This arrangement has shown to be successful in predicting the shear viscosity of supercritical $CO_2$ with thickening polymers [61]. Poiseuille flow is obtained in this work when a constant external force $Fp = 0.02$ (in reduced units) is applied along the $x$ – direction to all beads in the system. The beads that are within a distance $z \leq 0.15r_C$ of each wall are assigned zero velocity along the $z$ direction, to avoid slip [62 – 64]. Since the walls are placed perpendicularly to $z$ axis, periodic boundary conditions are applied only in the $x$ – and $y$ – directions. When the stationary state is reached, this setup produces a parabolic velocity profile along the $z$ – direction, as is shown in Fig. 2(b). The maximum shear rates in our simulations ($\dot{\gamma} = -(dv_x(z))/dz$), as extracted from the velocity profiles $v_x(z)$ in Fig. 2(b) go from $\dot{\gamma} = 0.07$, up to $\dot{\gamma} = 0.1$, in reduced units. It has been shown [65] that the typical Weissenberg numbers (We) for polymers with comparable polymer weight as those studied here under the shear rates modeled in our simulations are of order We~1. Therefore, the polymers have relaxation times for the extremal shear rates in our simulations that go from (in reduced units) λ~14.3 for the minimum shear rate, to λ~10.0 for the maximum shear rate. It is to be concluded that the relaxation times are of the same order for all polymers in the parameter range we have explored in this research. The parabolic velocity profiles presented in Fig. 2(b) are typical examples from our simulations and are characteristic of Poiseuille flow [54].

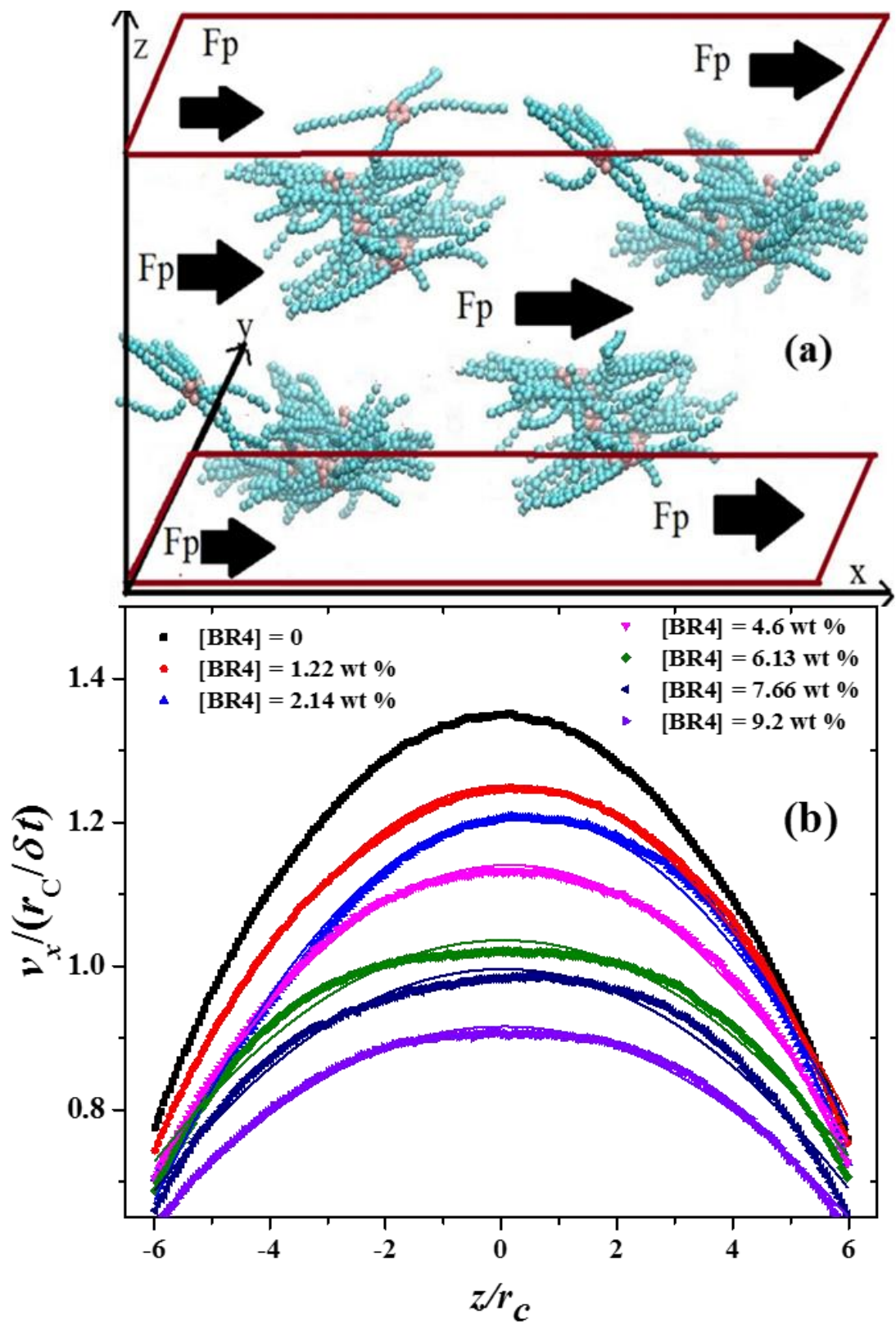

**Figure 2.** (a) Simulation setup used to obtain Poiseuille flow in this work. Cyan and pink beads make up the BR$_4$ copolymers with $N = 10$ beads along the branches. The solvent is not shown for clarity. The external constant force along the $x$ – axis $F_P(x) = 0.02\ k_BT/r_C$ is indicated by the arrows and imparts a velocity gradient to all the beads in the fluid. (b) Profiles of the $x-$ component of the fluid's beads velocity along the $z-$ axis for different concentrations of the thickening BR$_4$ copolymer. For all cases with BR$_4$ the bonding spring constant is $k = 100$, see Eq. (3), and the angle bonding constant is $k_\theta = 50$, see Eq. (4). The lines are the best fits to Eq. (8).

The shear viscosity $\eta$, was obtained after fitting the velocity profiles with the steady state solution for the Navier – Stokes equation, whose exact time-dependent solution is given by [54]:

$$v_x(z,t) = \frac{|F_p|}{2\nu}(z^2 - d^2) + \sum_{n=0}^{\infty} \frac{16(-1)^n d^2 |F_p|}{\nu\pi^3(2n+1)^3} \cos\left[\frac{(2n+1)\pi z}{2d}\right] \exp\left[-\frac{(2n+1)^2\pi^2\nu t}{4d^2}\right]. \quad (5)$$

In Eq. (5) $d$ is the half separation between the surfaces in Fig. 2(a), $\nu = \eta/\rho$ is the coefficient of kinematic viscosity, $\rho$ the number density and $F_p$ is an external force which is applied at each particle in the $x-$ direction, given by

$$|F_p| = \frac{1}{\rho}\frac{\partial p}{\partial x} = \frac{1}{\rho}\left(\frac{\Delta p}{L}\right) = -\frac{2\nu v_0}{d^2} = -\frac{2\eta v_0}{\rho d^2}. \quad (6)$$

In Eq. (6) $\Delta p$ is the pressure difference between two points separated by a length $L$ along the $x-$ direction and $v_0$ is the maximum asymptotic velocity defined by:

$$v_0 = -\frac{d^2}{2\rho\nu}\left(\frac{\Delta p}{L}\right). \quad (7)$$

When $t \to \infty$ the steady state is reached and the series expansion in Eq. (5) vanishes. The velocity profile approaches the steady-state solution [54]:

$$v_x(z) = \frac{|F|}{2\nu}(z^2 - d^2) = v_0\left(1 - \frac{z^2}{d^2}\right), \quad (8)$$

where $v_0$ is given by Eq. (7). By fitting the velocity profile obtained in the simulation with Eq. (8) it is possible to obtain $v_0$ and $v_0/d^2$; using then Eq. (6) one can calculate the coefficient of kinematic viscosity $\nu$ and then extract the viscosity, $\eta$.

## RESULTS AND DISCUSSION

In this section we present the results for three different case studies. The first one is devoted to exploring the effect in the viscosity of increasing the size of the branches, *N*, of the BR$_4$ prototype copolymer defined in the previous section (Fig. 1). The second case consists of studying the effect in the viscosity of varying the angular flexibility of the branches in the BR$_4$ copolymer by modifying the $k_\theta$ constant, see Eq. (4). In the last case study, the influence in the viscosity of increasing the strength of the bonding interactions through the spring constant $k$ (see Eq. (3)) is presented. In all cases, the viscosity is predicted at different copolymer concentrations in the solvent.

Before presenting the results corresponding to each case study, two tests of the choice of interaction parameters were carried out. The first, displayed in Fig. 3, shows the viscosity of the fluid as a function of polymer concentration for three values of the wall force intensity parameter, $a_w$, and for two values of the angular harmonic potential constant, $k_\theta$.

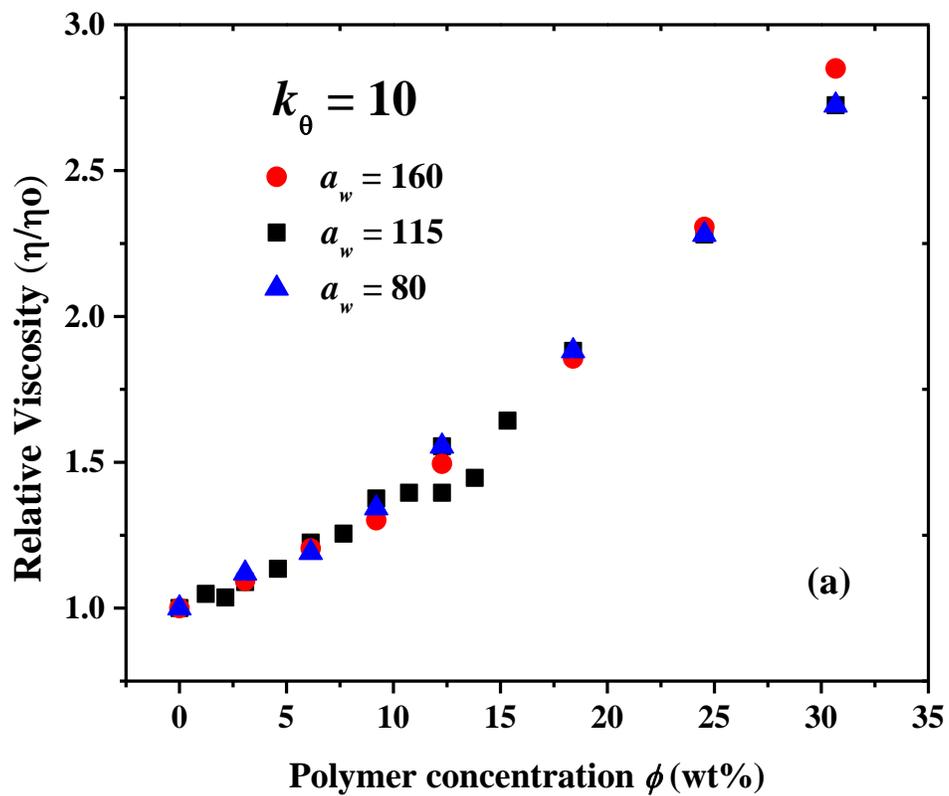

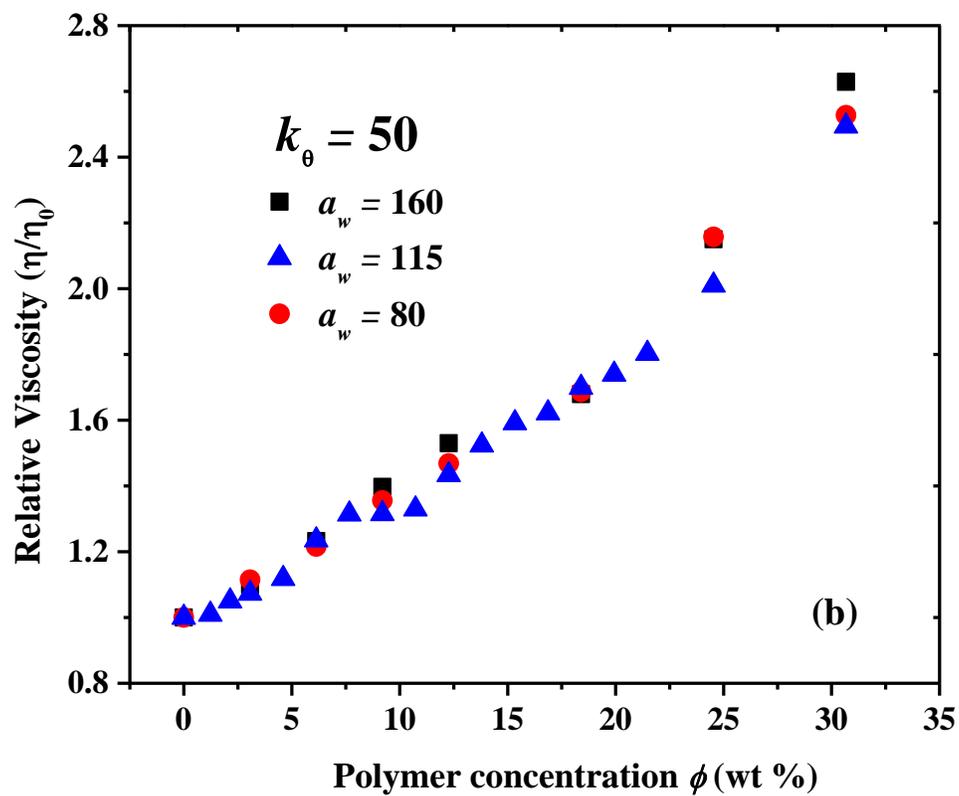

**Fig. 3**. Relative viscosity as a function of polymer concentration for different wall/fluid interactions, $a_w$, for two values of the angular spring constant: (a) $k_\theta = 10$ and (b) $k_\theta = 50$. In all cases the harmonic bonding interaction is $k = 100.0$.

The solvent/wall interactions were chosen identical to the polymer/wall interactions. The results presented in Fig. 3 indicate that the influence of the wall force intensity on the viscosity is minimal, confirming that the role played by the effective walls is that of providing only the boundary conditions necessary to confine the fluid. However, if the polymer molecules in the fluid experience attraction toward the walls it may lead to adsorption of those molecules, which in turn may affect the viscosity of the confined fluid. In this work, the polymer – wall interactions are the same as the solvent – wall interaction, thus no polymer adsorption on the surfaces is observed. The second test corresponds to the choice of the conservative force, particle – particle interaction parameters, see Eq. (2). Simulations were performed for two solvent conditions: good solvent and theta solvent. Three cases were studied for each of those solvents, namely for angular potential constants equal to $k_\theta = 10, 50$ and $100$ (in reduced units). The results can be found in Fig. 4 and show that the viscosity is larger for a good solvent, as expected, although very little influence is found on the angular potential constant.

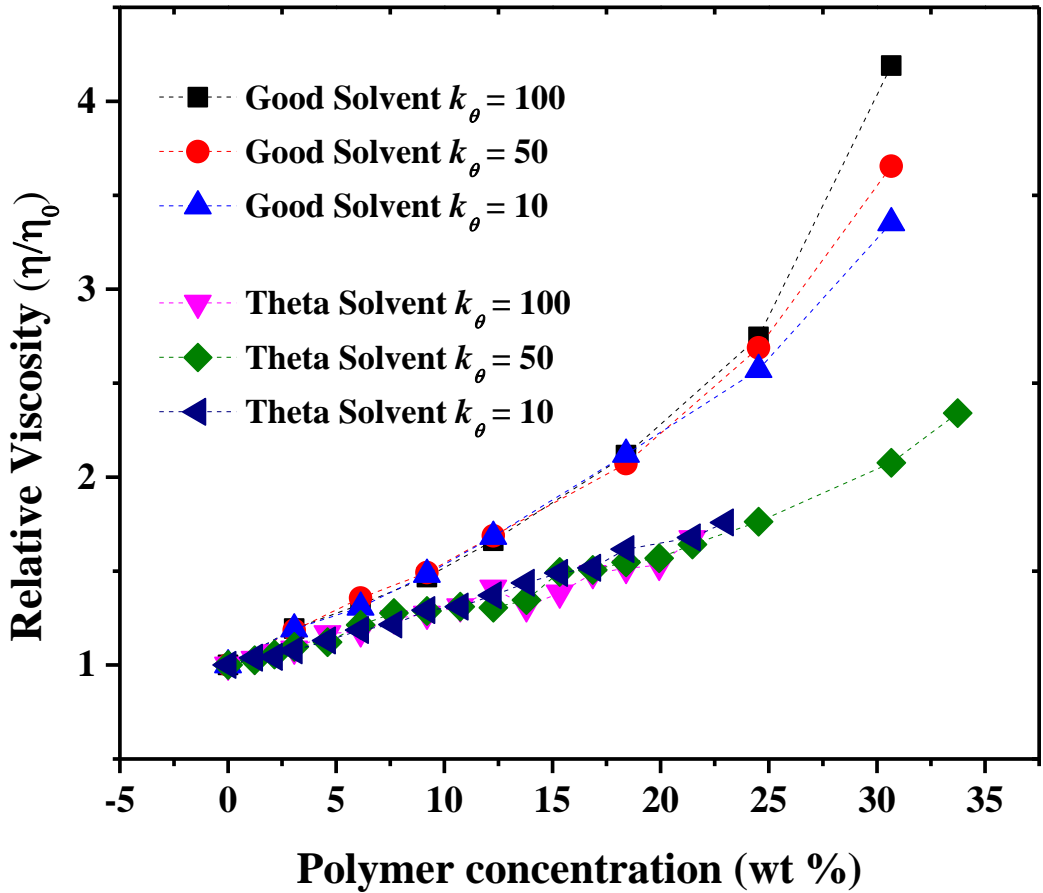

**Fig. 4**. Relative viscosity as a function of polymer concentration for different DPD interaction parameters, $a_{ij}$, to model theta, and good solvent conditions. For each solvent quality, simulations were performed for three values of the constant of the angular harmonic potential between three neighboring beads along the polymer chain. See the text for the values of the DPD interaction parameters chosen for each solvent.

The conservative force interaction parameters used for the predictions shown in Fig. 4 are listed in Tables 1 and 2, corresponding to theta-solvent and good-solvent conditions, respectively. To check that the polymers are small compared with the channel length, so that the local shear can be considered as constant, the radius of gyration was calculated. Additional simulations were carried out to calculate it for polymers confined in a channel of width $L_z^* = 17.0$ in reduced DPD units, with bonding interaction constant $k = 100.0$ in reduced DPD units. There were 80 polymer molecules in the simulation box, corresponding to a concentration of 24.53 wt %. These new simulations predicted the radius of gyration to be $R_G^* = 11 \pm 1$, in reduced DPD units. Therefore, the polymers can be considered as small compared with the channel width.

**Table 2**. DPD interaction parameters ($a_{ij}$) between the solvent beads (S), the beads representing the benzene rings (B) and the branches (R) of the polymers under good solvent conditions. See Eq. (2).

| $a_{ij}$ | S | B | R |
|---|---|---|---|
| S | 78.3 | 120 | 60 |
| B |  | 50 | 100 |
| R |  |  | 79.3 |

The structure of the fluid made up of solvent beads and the copolymers was studied also in the bulk, i.e. unconfined and in equilibrium, confined by walls but in equilibrium and lastly, confined and under Poiseuille flow. The resulting density profiles are shown in Fig. 5(a) for the branches (R), and in Fig. 5(b) for the central structures (B).

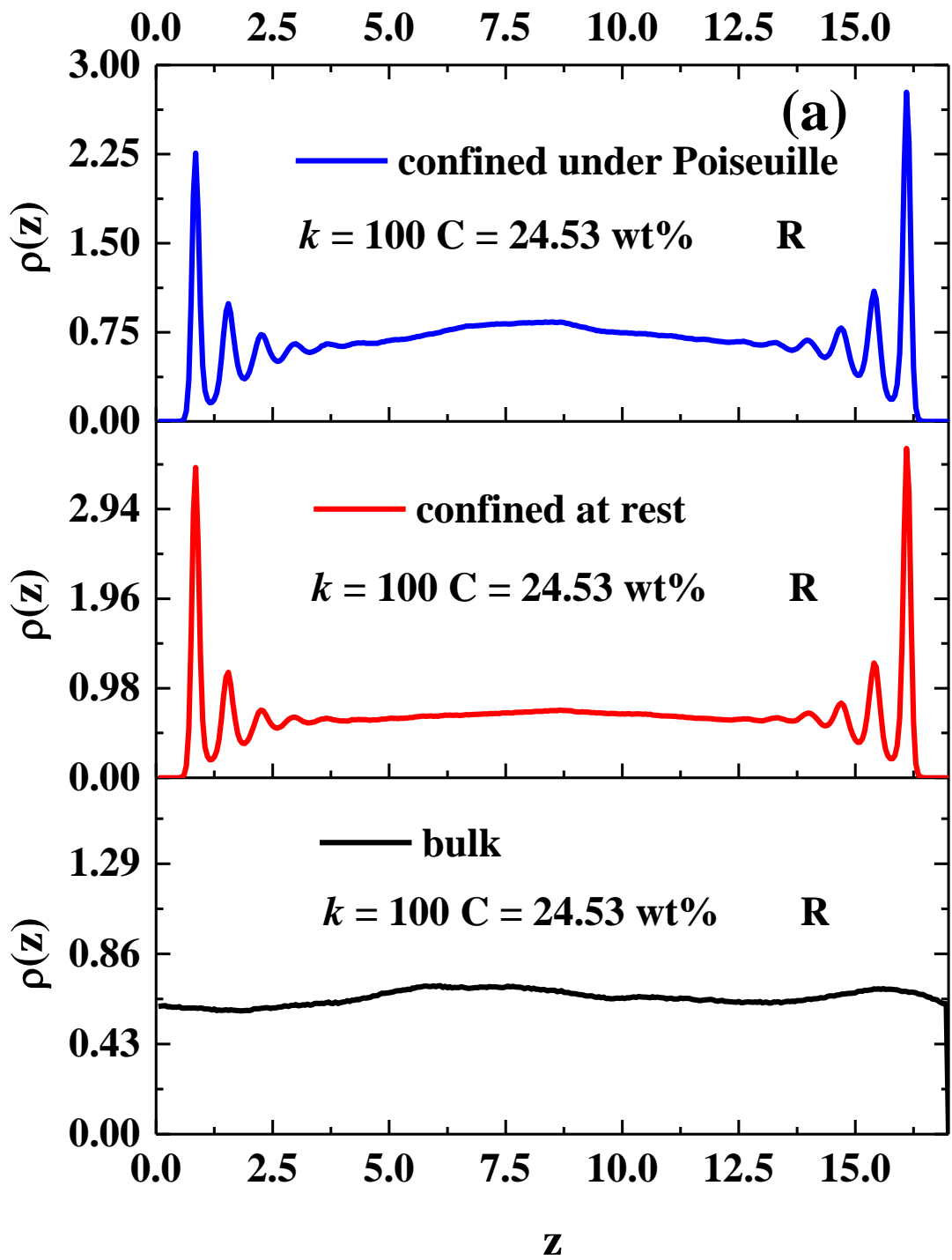

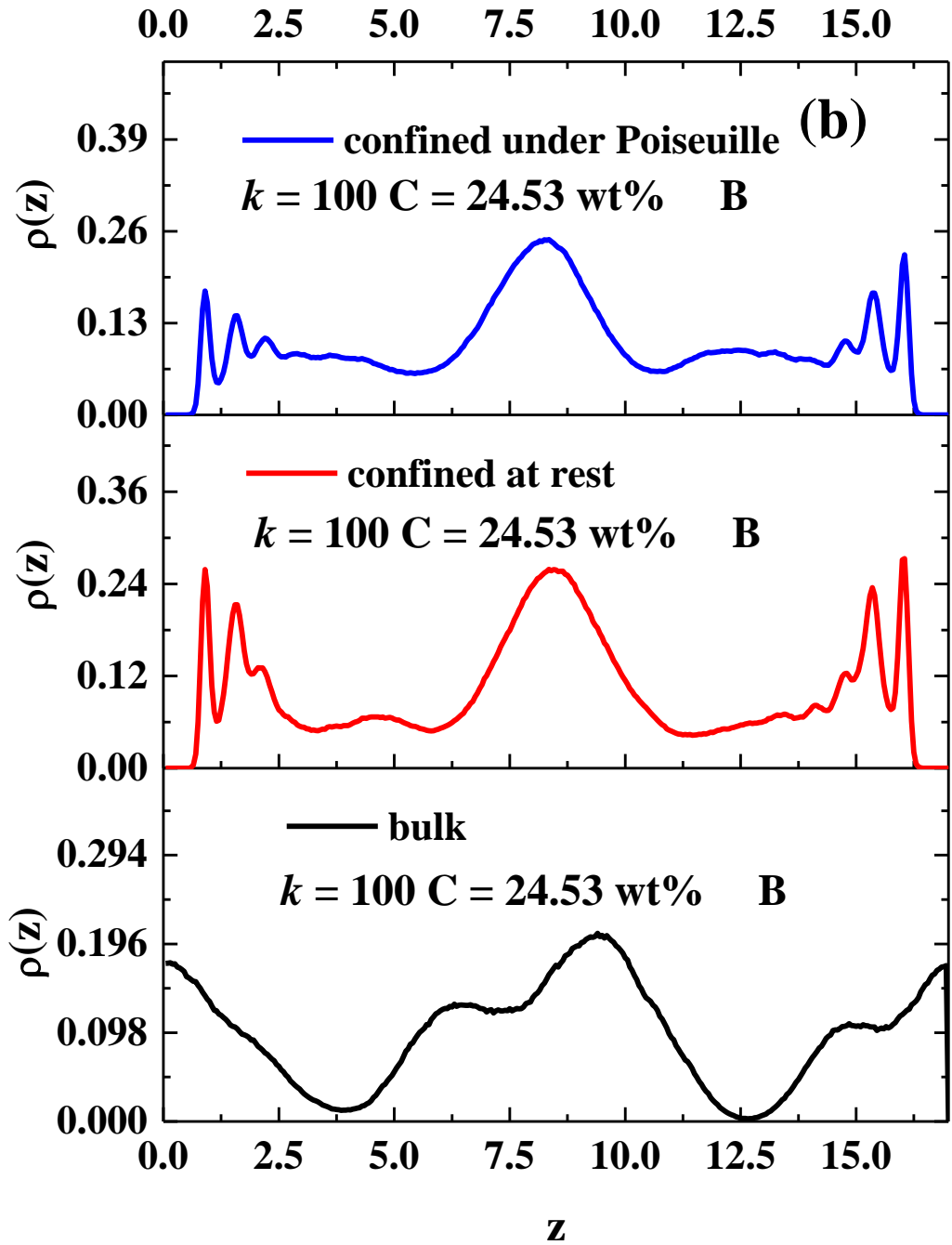

**Fig. 5**. (a) Density profiles of the beads that make up the branches of the copolymers (R), and (b) density profiles of the central beads in the copolymers (B). The profiles were obtained from bulk simulations (black lines), confined fluid in equilibrium (red lines), and confined fluid under Poiseuille flow (blue lines). In all cases the branches (R) have 10 beads, the copolymer concentration is 24.53 wt %, the bonding constant is $k = 100.0$ (see Eq. (3)) and the angular constant is $k_\theta = 50.0$ (see Eq. (4)), both in reduced units.

The profiles displayed in Fig. 5 show that the density in the center of the channel is approximately constant, that the profiles are symmetric and that their structuring (appearance of maxima and minima near the walls) is a consequence of the confinement only. The structuring is always found in confined systems because the particles tend to organize in layers near a surface, regardless their interactions [39]. The flow does not affect the structure of the fluid as much as the confinement.

**Influence of the branches' length**

The length ($N$) of the arms or branches (labelled R) of the star polymer BR$_4$ was modeled for $N = 1, 5, 10$ and $20$ DPD beads. Fixed values of $k = 100$ (see Eq. (3)) and $k_\theta = 50$ (Eq. (4)) were used in all of these cases. The resulting aggregates formed by the intermolecular association of the copolymers of different sizes $N$ of branches R in the solvent S are presented in the snapshots in the left and center panels in Fig. 6. In all the cases shown in Fig. 6 there are 20 molecules of the BR$_4$ copolymer, with branch lengths equal to $N = 1, 5, 10, 20$ as indicated by the labels in the figure.

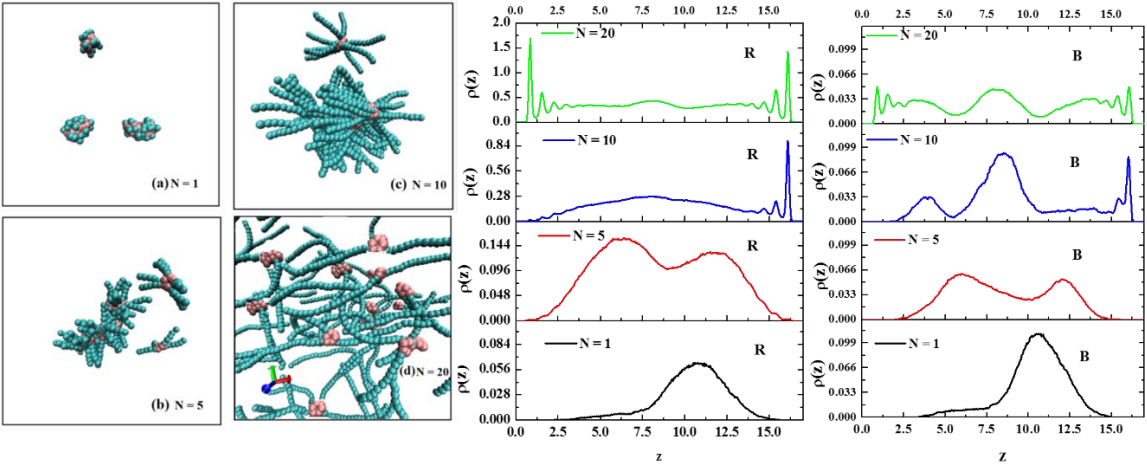

**Figure 6**. (Two leftmost panels) Snapshots of the polymer associations for different branch lengths in BR$_4$. Pink beads correspond with the central structure of the copolymers (labelled as B in Fig. 1) while cyan beads correspond to the linear branches (labeled as R in Fig. 1). In

(a) there are 20 molecules of the BR$_4$ copolymer whose branches have length $N = 1$; in (b) there are 20 BR$_4$ molecules with $N = 5$. In (c) and (d) there are also 20 BR$_4$ copolymers with length branches $N = 10$ and $N = 20$, respectively. In all cases, $k = 100.0$ and $k_\theta = 50.0$, both in reduced units. The solvent is omitted in all snapshots, for simplicity. Density profiles of the beads in the central structure of the copolymers, B (rightmost panel), and density profiles of the beads in the branches of the copolymer, R (second rightmost panel), for $N = 1, 5, 10$ and 20 beads.

The snapshots presented in Fig. 6 for different lengths of the arms of the star – like BR$_4$ copolymer show that aggregates are formed, which become more intertwined as $N$ grows. This grouping is driven by the intermolecular attraction of the beads that make up the central structure of the copolymers, B. When the branches are short ($N = 1, 5$) there is more space available for the B beads to come closer to one another, forming micelle – like aggregates. Little viscosification increase is expected in these cases, since there is small momentum transfer for relatively small structures. As $N$ grows, steric interactions lead to the formation of networks, see the snapshots for $N = 10, 20$ in Fig. 6. The density profiles of the B beads in the central structure of the copolymers (rightmost panel in Fig. 6) confirm that they are responsible for the intermolecular association. That is because the profiles show maxima in the central region of the box. These beads play the role of the aromatic rings in the fluorocarbon – based copolymers used to viscosify supercritical $CO_2$ [5]. The density profiles of the R beads (second rightmost panel in Fig. 6), in the branches of the copolymers, show they are uniformly distributed along the $z$ – axis of the simulation box. This is a consequence of the good solvent conditions under which those beads are, which explains their extended configurations in the snapshots shown in Fig. 6.

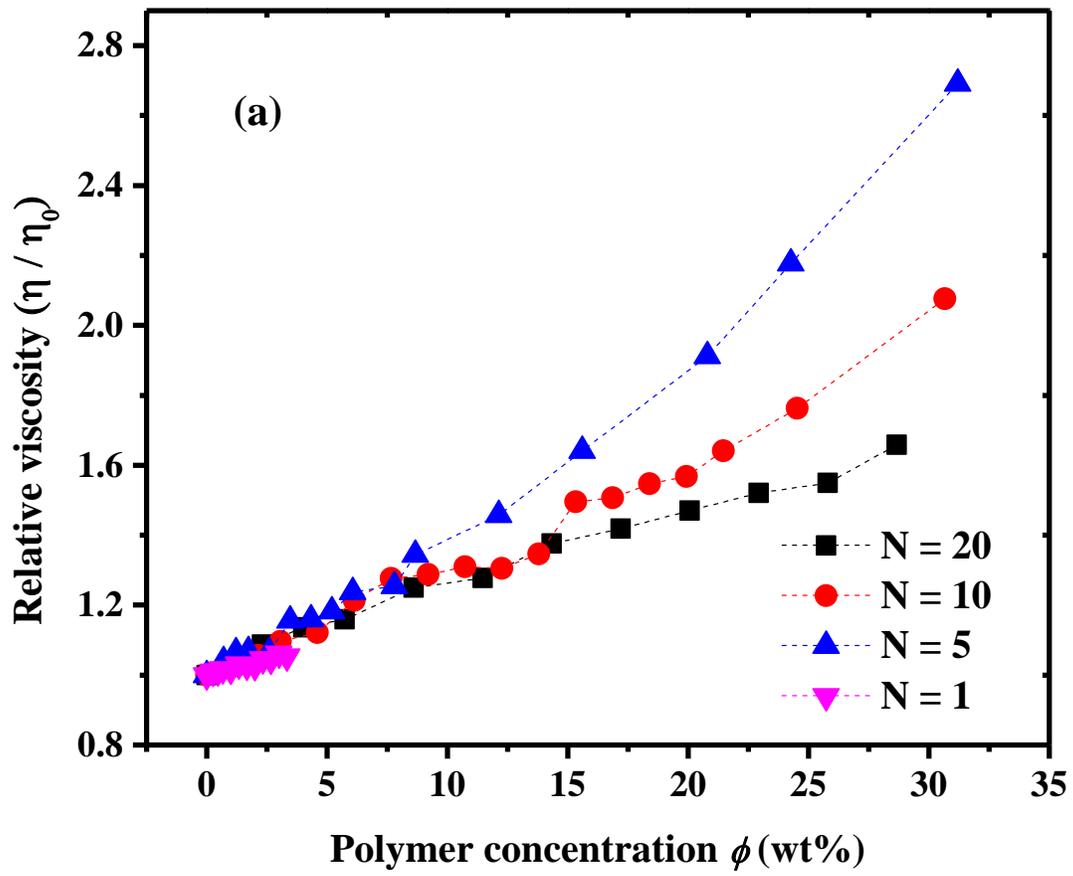

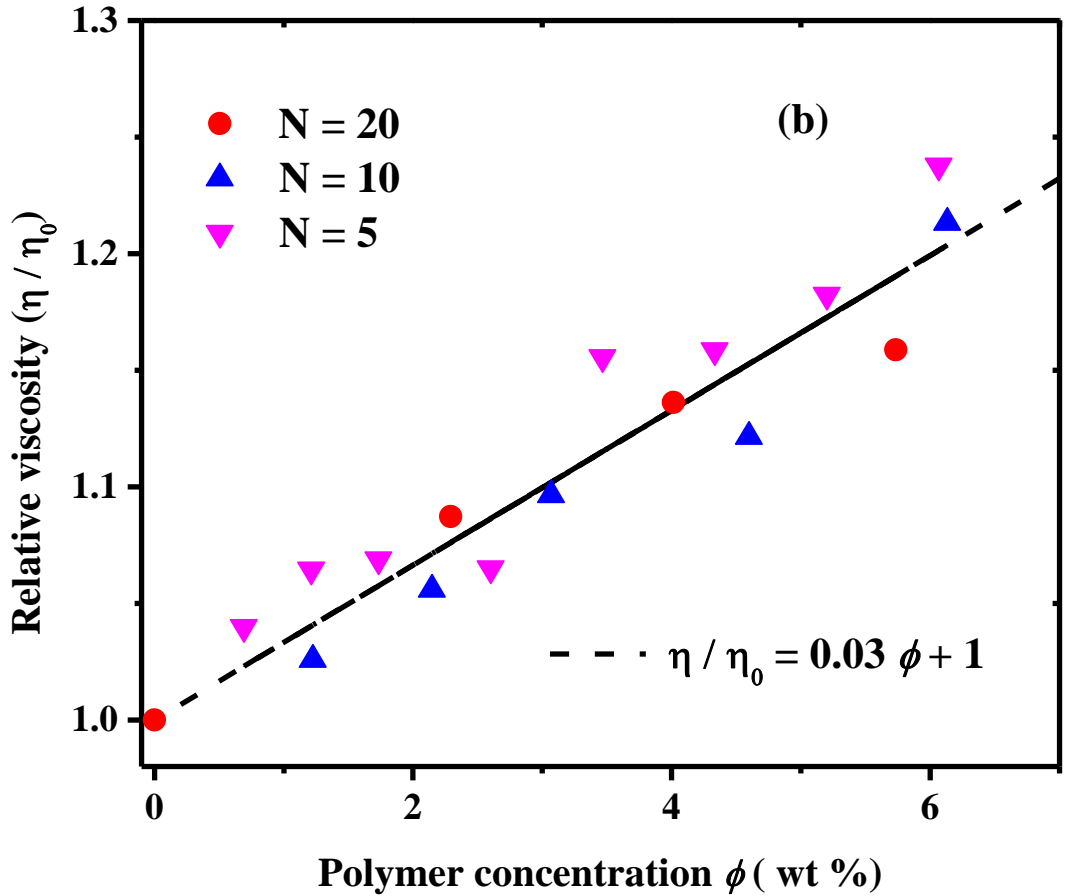

**Figure 7**. (a) Relative viscosity ($\eta/\eta_0$) as a function of copolymer concentration for increasing values of the polymerization degree of the branches (N). (b) Linear fit of the relative viscosity as a function of concentration for the dilute region $\phi < 7$ for $N = 20$, 10 and 5, with $k = 100$ (see Eq. (3)) and $k_\theta = 50$ (see Eq. (4)).

Relative viscosity results are shown in Fig. 7, where $\eta/\eta_0$ is presented as a function of the concentration $\phi$ expressed in weight percentage (wt %), $\phi = (N_p M/N_T)(100)$ where $N_p$ is the number of polymers added, $N_T$ is the total particles in the simulation and $M$ the reduced molecular weight. As expected, increasing the concentration of copolymer, in the cases where the branches are present (N $\neq$ 1), an increase in the viscosity of the solvent is achieved. The increase in viscosity reached is around 20 % in dilute regimen at $\phi = 5$ wt %, see Fig. 7(a). For the semi-dilute ($0 < \phi^* = \phi/\phi_{max} \leq 0.1$) region there is an increase of 40 % for $\phi = 10$ wt %, and at high concentrations we obtain better increase approximately of 170 % in viscosity for $\phi = 30$ wt % for the best thickener corresponding with $N = 5$. Analysis in the diluted region ($\phi < 10$ wt %) shows linear behavior for the viscosity as a function of concentration $\phi$, as is shown in Fig. 7(b), in agreement with Einstein's prediction [66]. For the concentrated regime ($\phi > 10$ wt %, $0.5 \sim \phi^* < 1$), see Fig. 8, the relative viscosity follows a power law as a function of concentration, given by:

$$\frac{\eta}{\eta_0} = A\,\phi^{\alpha} \ . \tag{9}$$

The predictions shown in Fig. 8 can be fitted to Eq. (9), which is known as the Mark – Houwink equation [66]. The constants $\alpha$ and $A$ are fitting parameters which depend on the specific polymer – solvent system, with $\alpha = 0.5$ for theta solvent and $\alpha = 0.8$ for good solvents [66]. They also depend on the polymers' persistence length: for flexible polymers, the exponent $\alpha$ in Eq. (9) is found in the range $0.5 < \alpha < 0.8$; for semi-flexible polymers, and $\alpha \geq 0.8$ for rigid polymers $\alpha \geq 2$ [67]. As the fits to the data in Fig. 8 show, we predict values of $\alpha$ in Eq. (9) that are in the range $0.43 \leq \alpha \leq 0.68$, which means the polymers are flexible and are under theta solvent conditions. The influence of increasing the molecular weight of these star – like polymers through increments of the polymerization degree ($N$) of the branches does not appear to be an effective method to raise the solvent's viscosity at low polymer concentration.

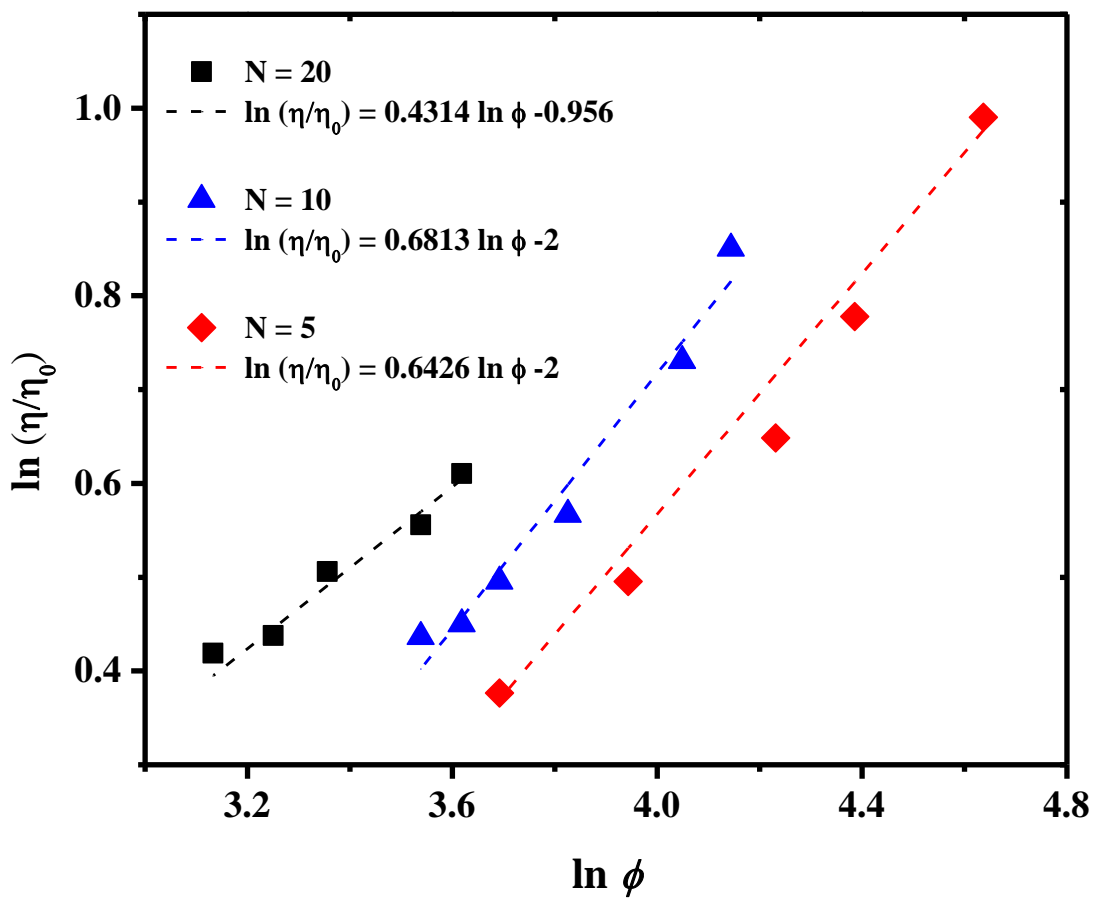

**Figure 8.** Relative viscosity as a function of copolymer concentration (in reduced units) for increasing length of the branches of the BR$_4$ copolymers. The dashed lines represent the best fits to Eq. (9); the $\alpha$ exponents are the slopes of the fits.

**Influence of the flexibility of the branches**

The effect of the flexibility along the branches of the star – like copolymer in the viscosity was studied by modifying the constant $k_\theta$ = 10, 50 and 100, see Eq. (4). Snapshots of the resulting structures formed by the copolymers, and their density profiles are presented in Fig. 9 for 40 molecules of BR$_4$ (12.26 wt %) with $N$ = 10 and $k$ = 100. The copolymers aggregate more easily when their linear branches are flexible (see top left panel in Fig. 9, for $k_\theta$ = 10) because the B beads on different molecules attract each other. Steric hindrance is reduced when the branches are very flexible. The density profile of the B type beads in the blue line shown in bottom left panel in Fig. 9 shows three major maxima, meaning that there are three major aggregates of the BR$_4$ copolymers. Reducing the flexibility of the copolymers arms, i.e. increasing $k_\theta$ modifies the structure of the fluid. The copolymers do not aggregate as easily as they do when their branches are flexible, but they still form aggregates, as shown by the red line ($k_\theta = 50$) in the B – type beads density profile in the bottom left panel in Fig 9. When the thickening copolymers have the stiffest branches (top right panel in Fig. 9) their associations occur mostly through the overlap of their branches. The density profiles of the branches, labelled R and shown in the bottom right panel of Fig. 9 confirm that they are uniformly distributed between the walls.

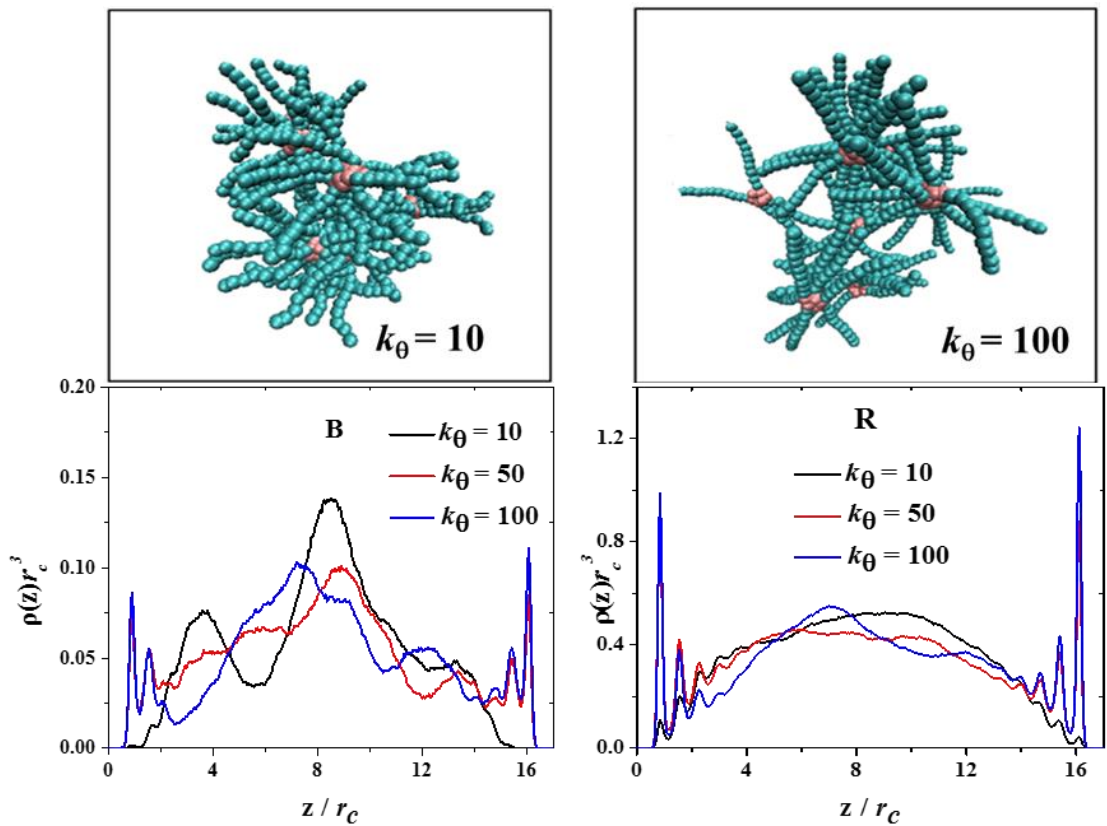

**Figure 9**. (Top panels) Snapshots of the association of 40 BR$_4$ copolymers (12.26 wt %) with $k_\theta = 10, 100$. Pink beads correspond to the B beads of the structure while cyan beads represent the linear polymeric branches R, see also Fig. 1. (Bottom panels) Density profiles of B and R beads in the copolymers for three values of the $k_\theta$ constant, all in reduced units.

The influence of the modification of the copolymers' branches flexibility on the viscosity is presented in Fig. 10, as a function of copolymer concentration. The flexibility of the copolymers' branches is found to have less influence on the viscosity of the fluid than, for example, the branches length (see Fig. 7(a)). The viscosity increases linearly with copolymer concentration, following Einstein's model [66, 68], with a slope similar to that found when the chains' length is increased (see Fig. 7(b)). Figure 10 shows that the viscosity of the fluid is basically set by the copolymer concentration and the length of its branches, with little influence of the branches' stiffness.

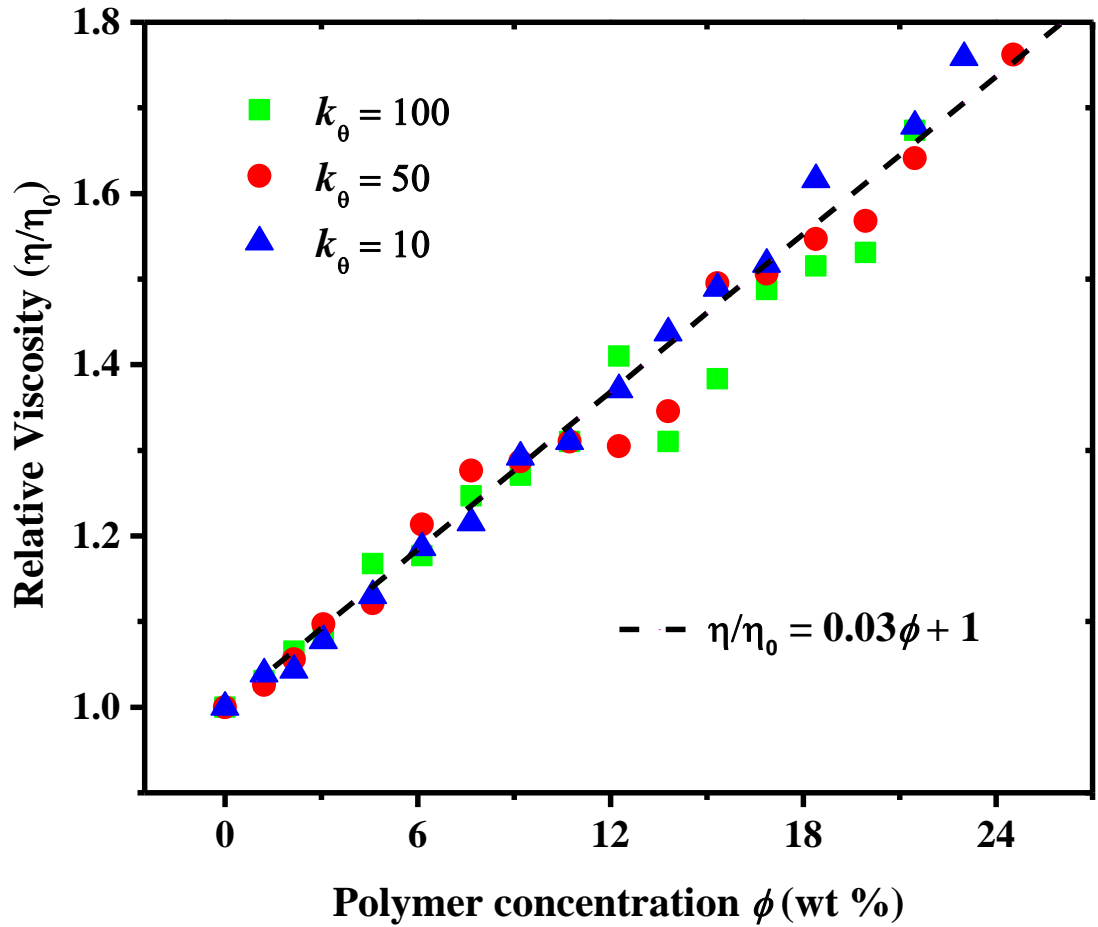

**Figure 10.** Relative viscosity $\eta/\eta_0$ of the fluid with BR$_4$ copolymers as a function of the BR$_4$ concentration, with $k_\theta = 10$, 50 and 100, see Eq. (4); with $k = 100$, see Eq. (3), and $N = 10$ (polymerization degree of the branches). All $k_\theta$ values are reported in reduced units. The dashed line is the best linear fit. See text for details.

**Influence of the bonding interactions**

Finally, we present the results for the effect in the relative viscosity originated by modifying the harmonic constant *k,* see Eq. (3), which represents the intensity of the bonding interactions between the DPD beads that make up the thickening copolymers. Simulations were performed for the following *k* values (in reduced DPD units): $k = 4, 5, 10, 40, 50$ and $100$. Only the constants corresponding to the bonds between the beads along the branches of the copolymers were varied, called R in Fig. 1(a). The angular bonding constant was chosen as $k_\theta = 50$ in all cases. The bonds holding the B – type beads in the center of the BR$_4$ structure remain always constant, see Fig. 1(a).

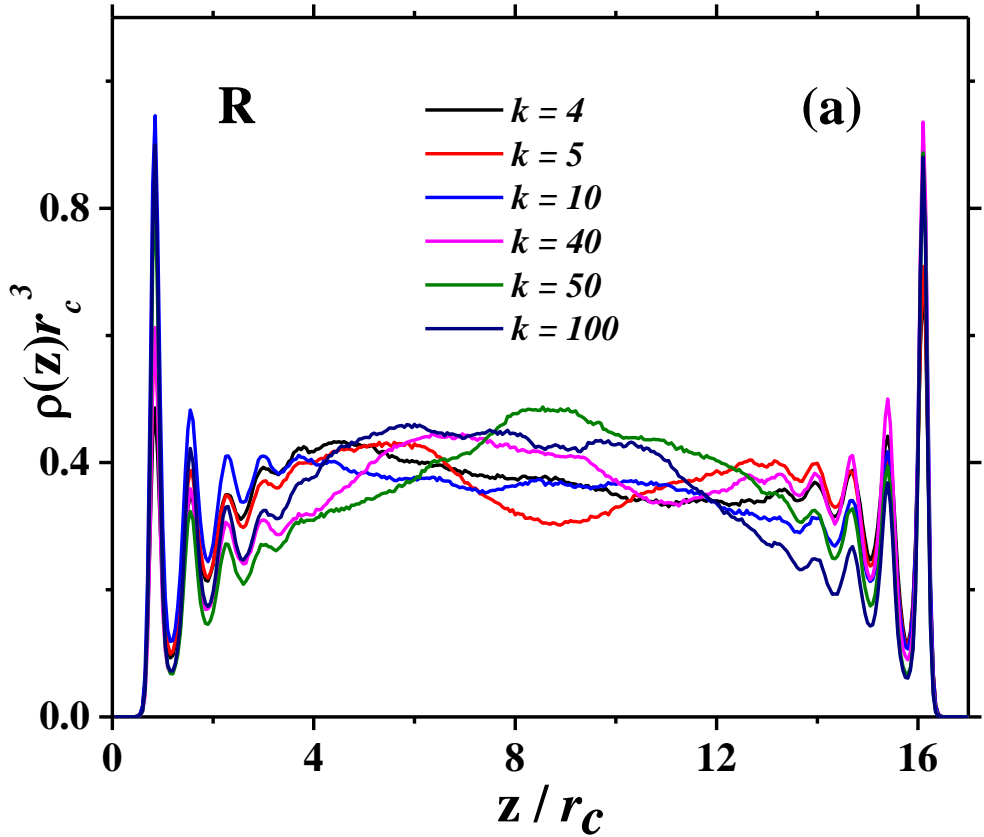

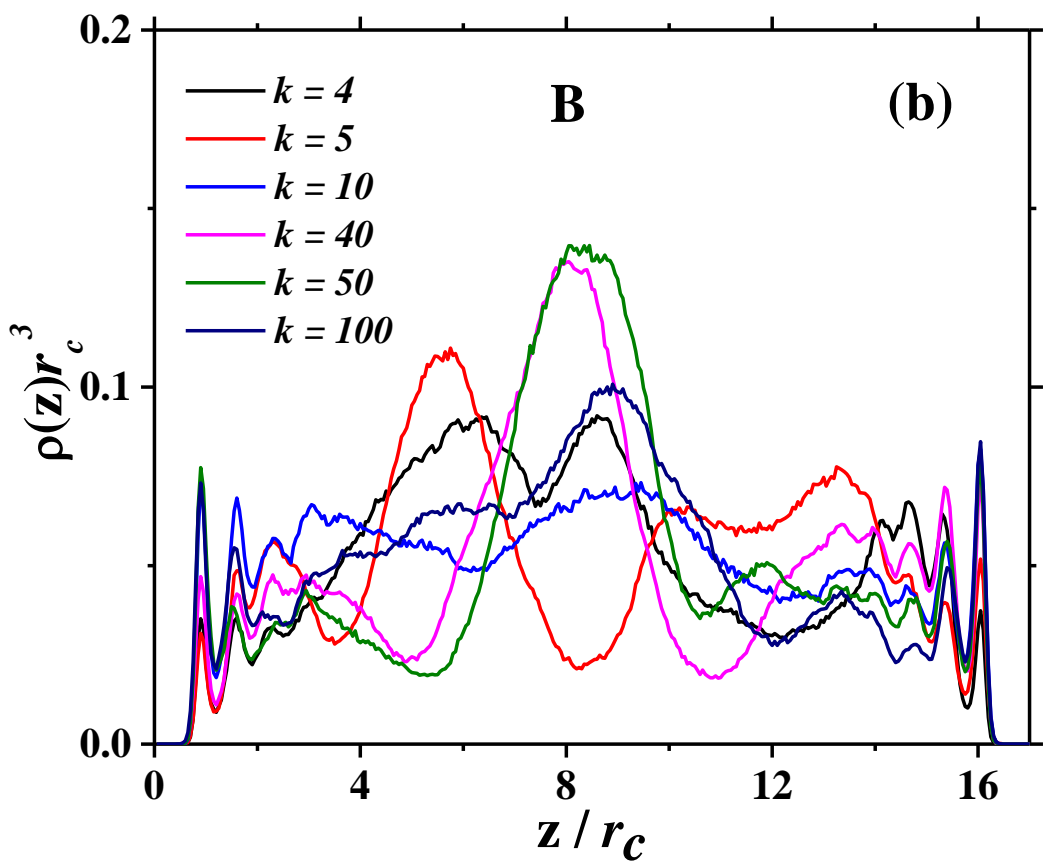

**Figure 11**. Density profiles for the beads labeled (a) R and (b) B of the BR$_4$ copolymers, see also Fig. 1. There are 40 BR$_4$ molecules (12.26 wt %) in the system, all with branches whose polymerization degree is $N = 10$. All $k$ values are expressed in reduced DPD units. See text for details. The angular constant is $k_\theta = 50.0$ in reduced units, for all cases.

A small $k$ value makes the formation of intermolecular aggregates more effective, that is, there is a greater distribution of aggregates throughout the system and the branches can form larger networks. This is confirmed by the density profiles shown in Fig. 11, where the profiles for the copolymer center beads labelled B display three maxima for the smallest $k$ values. As the spring constant is increased the B – type beads tend to migrate to the walls, leading to the strong layering seen in the density profile when $k = 100$, for example. The profiles for the copolymer branches, labelled R, are distributed more or less uniformly between the walls, but when the spring constant value increases beyond $k = 10$ the profiles for the branch beads show a maximum at the center of the pore between the walls. The stronger bonded beads make the copolymers become shorter and their adsorption on the walls and intermolecular associations are reduced, which are detrimental for the viscosification of the fluid.

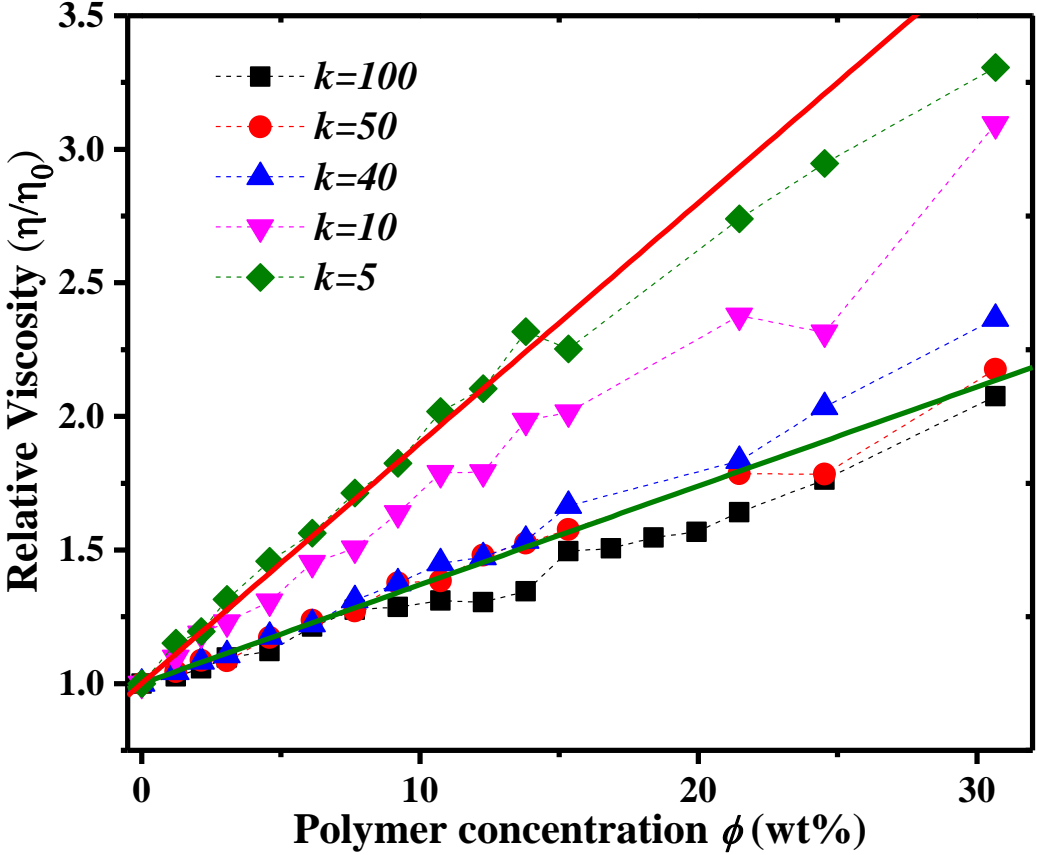

**Figure 12.** Relative viscosity of the copolymer/fluid system as a function of concentration for BR$_4$ copolymers with bonding constant interactions equal to $k = 5, 10, 40, 50, 100$, in reduced DPD units. The branches polymerization degree of the branches is $N = 10$. The solid lines represent linear fits with slopes equal to 0.037 (green line) and 0.09 (red line).

The relative viscosity increase obtained when the bonding interactions are varied is presented in Fig. 12. There are roughly two regimes as the bonding interaction strength is increased. In the first regime for bonding spring constant values (in reduced DPD units) equal to $k = 40$, 50 and 100 (considered as strong bonding) one finds roughly the same slope in the viscosity:

0.037. Notice this value is similar to the slopes obtained in the relative viscosity when the polymerization degree of the BR$_4$ molecules is increased (Fig. 7(b)) and when the angular interaction is varied (Fig. 10). In the dilute regime, $\phi$ (wt%) $\leq 8$, and for strong bonding ($k = $ 40, 50 and 100) the increase in the viscosity is similar to the previous cases, around 20% for $\phi = 5$ wt%, 40 % for $\phi = 10$ wt %, and at high concentrations the increase reaches up to 120 % in viscosity for $\phi = 30$ wt %. For the cases where the bond is weaker ($k = 5$, 10, in reduced DPD units) a better performance in the viscosity is found, with an increase of around 50 % for $\phi = 5$ wt% and 90 % for $\phi = 10$ wt % when $k = 5$. At high concentrations the increase in viscosity is 230 % for $\phi = 30$ wt %, also when $k = 5$. The different viscosification potential obtained while varying the bonding interaction of beads along the branches of the BR$_4$ copolymers can be understood from the density profiles in Fig. 11. When the branches are very flexible ($k = 5, 10$, in reduced units) they extend uniformly between the walls, see the snapshot for $k = 10$ in Fig. 11. The same trend is found in the density profile of the branches, labeled R in Fig. 11, where the density profiles corresponding to $k = 5$ and 10 are uniformly distributed in the center of the pore formed by the walls. This is in contrast to the density profiles shown in the same figure for $k = 40, 50$ and 100, all of which have a maximum in the center of the pore. Softer bonding interactions allow the branches to stretch and intermingle with branches from other copolymers, creating agglomerates that yield larger viscosity [52].

Exploring the dependence of the viscosity with the spring constant, universal $k$ – independent behavior is obtained when the $\eta/\eta_0$ data from Fig. 13 are plotted as a function of $\phi_r = \phi/\ln(k)$, see Fig. 13. The universal scaling function obtained is

$$\frac{\eta}{\eta_0} = 0.13\phi_r + 1.04 \qquad (10)$$

where the appearance of the spring constant of the bonding force between beads in the denominator shows that the largest increments in the viscosity occur for the weakest bonds. Therefore, the variable that most influences the increase in the viscosity out of those studied in this work is the strength of the harmonic spring constant in the force that models the bonding interactions between the beads that make up the branches of the $BR_4$ copolymers.

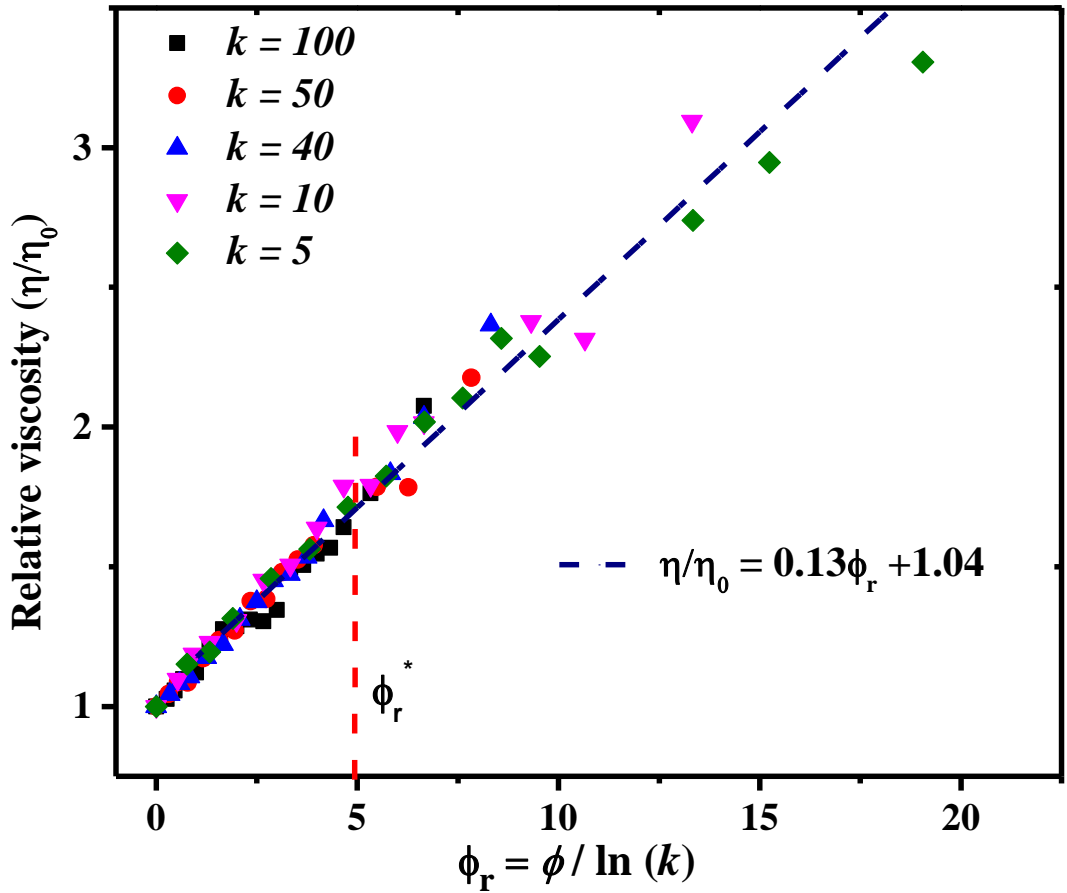

**Figure 13**. Universal scaling of the relative viscosity data shown in Fig. 12, as a function of the variable $\phi_r = \phi/\ln(k)$, where $k$ is the constant of the spring bonding beads along the branches of the thickening copolymers. The dashed red line defines $\phi_r^*$ and represents the linear regime ($0 \leq \phi_r^* \leq 5$) for the relative viscosity.

The relative viscosity of the fluid under Poiseuille flow beyond the linear regime, i.e. for $\phi_r^* > 5$ in Fig. 13 can be better represented by a power law rather than a linear relationship.

The data and the power – law fit are presented in Fig. 13. This rescaled universal plot for the viscosity as a function of $\phi_r = \phi/\ln(k)$ yields a scaling with respect to the dependence with concentration which is similar to that found when increasing the branches' length, see Eq. (9), namely:

$$\frac{\eta}{\eta_o} = A\phi_r^\alpha \qquad (11)$$

where $\alpha = 0.48$ and $A$ is fitting constant. Equation (11) is the Mark – Houwink equation [66] for the viscosity increase in a fluid produced by thickening polymers and the exponent obtained corresponds to the theta solvent case [67]:

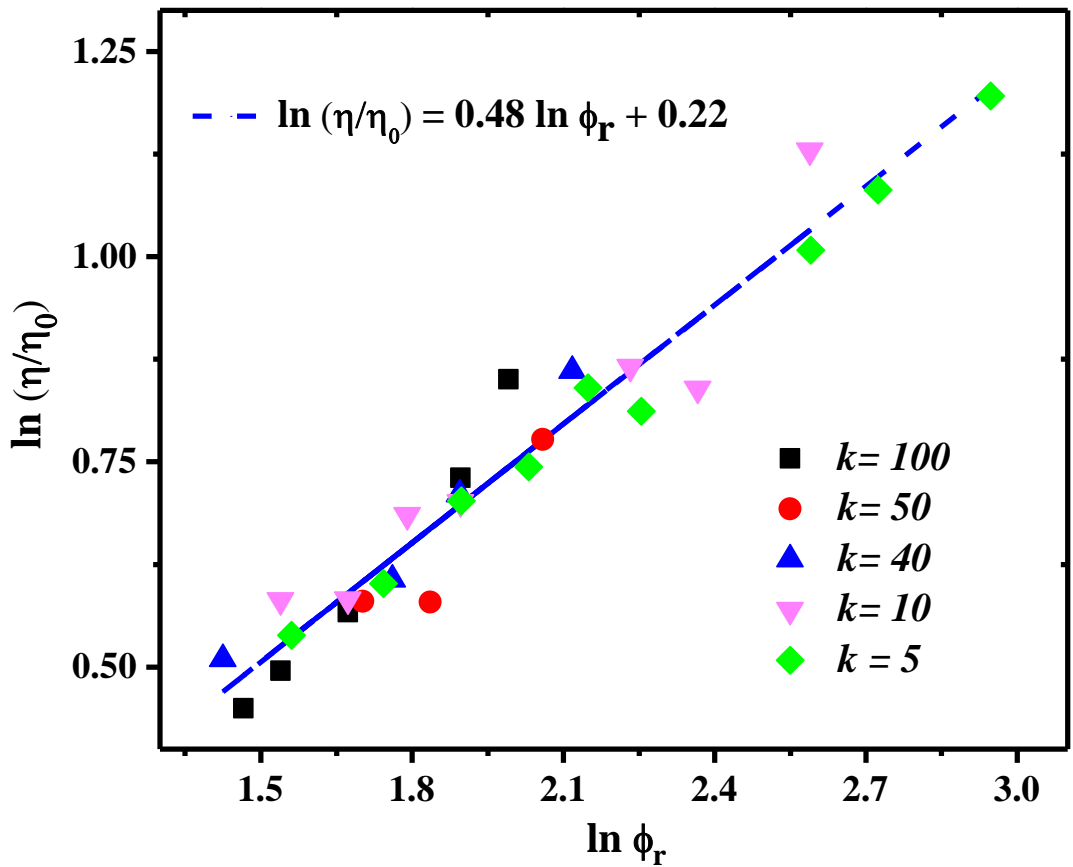

**Figure 14**. Scaling behavior of the logarithm of the relative viscosity as a function of the logarithm of $\phi_r = \phi/\ln(k)$, where $k$ is the bonding constant of the spring bonding beads along the branches of the BR$_4$ copolymers. The dashed line represents the power law best fit.

For larger copolymer concentrations the relative viscosity curves as functions of $\phi_r = \phi/\ln(k)$ all fall approximately on the same curve regardless the harmonic constant value, $k$, as shown in Fig. 14. The role of the $k$ constant value in the viscosity appears to be secondary when plotted as a function of $\phi_r = \phi/\ln(k)$ at high concentration since all five $k$ value curves follow roughly the power law given by Eq. (11). The latter equation and Fig. 14 show that the viscosity of the fluid with thickening polymers of different bonding strength between their beads still obeys the Mark–Houwink equation, with a renormalized polymer concentration, $\phi_r = \phi/\ln(k)$. A notable fact is that the exponent in Eq. (11) is $\alpha = 0.48$, regardless the $k$ value, corresponding to theta solvent conditions [67]. This suggests that a route to increase the viscosity at low polymer concentration consists of increasing the polymer solubility, regardless their bond strengths.

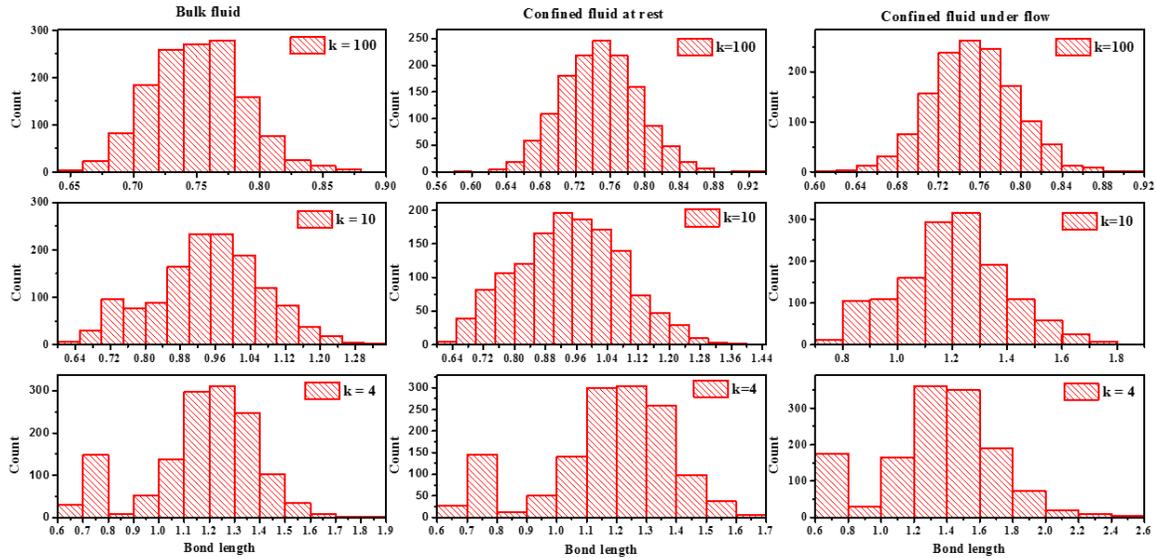

**Figure 15**. Histograms to the length of the bonds joining beads along the branches of the polymers, for three harmonic spring values (in reduced units): $k = 100$ (top row), $k = 10$ (middle row) and $k = 4$ (bottom row). There are 30 polymer molecules (9.2 wt %) in each simulation box. The leftmost column corresponds to simulations of the fluid in bulk (unconfined). The middle column is for confined fluids at rest (no flow). The rightmost column shows the histograms for the confined fluid under Poiseuille flow.

Lastly, we show in Fig. 15 the histograms of the bond lengths for 30 polymer molecules in the simulation box, under three conditions: in the bulk (unconfined); confined at rest and confined under Poiseuille flow. For each case, histograms were produced for three values of the spring's bond constant, namely $k = 100, k = 10$ and $k = 4$, in reduced units. The results presented in Fig. 15 show that for the stiffer bonds, those with $k = 100$ and $k = 10$, the histograms have a single maximum at a distance close to the equilibrium length, $r_0 = 1.0$ (in reduced units). This happens whether the fluid is in the bulk, confined at rest or confined under flow. For the weakest bond ($k = 4$) the histograms shown in Fig. 15 display two maxima, regardless of the boundary and flow conditions. One of them is found at $r \sim 0.75$ and the other at $r \sim 1.25$. Neither of these values represent a nonphysical bond stretching from the spring's equilibrium length, $r_0 = 1.0$.

## CONCLUSIONS

The viscosification of a model fluid with added thickening branched copolymers is predicted in this work under parabolic, Poiseuille flow using DPD mesoscale numerical simulations. Branched polymers were used because experiments [5, 10, 17] and simulations [53, 61] have shown them to be more effective in raising the viscosity than linear ones, under similar thermodynamic conditions. The viscosity is obtained as a function of polymer concentration, branch length and flexibility and as a function of angular bonding strength. Changing the latter is equivalent to varying the branches' persistence length. Out of all these variables, we find that the angular bonding strength ($k_\theta$) is the one that yields only linear increases in the viscosity, for all the concentrations studied here. This variable can be modified in experiments using, for example, polyelectrolytes at different pH [69]. The influence of the molecular weight of the branches (N) is also linear and with the same slope as that obtained when changing $k_\theta$, but only at relatively low copolymer concentration ($\phi \leq 6$ wt %). Also, it is found that increasing N yields lower viscosity increase, contrary to what would be expected for linear polymers. For large copolymer concentration ($\phi > 6$ wt %) the viscosity obeys the Mark –Houwink power law equation, where the exponent tends to be larger for smaller N. The most complex effect is obtained when the strength of the bonding interactions (*k*) is modified. For all concentrations the viscosity displays a linear dependence, albeit with a slope that depends on the bonding strength, being larger for smaller *k* values. Furthermore, we find it is possible to obtain $k-$ independent trends for the viscosity if the copolymer concentration ($\phi$) is renormalized as $\phi_r = \phi/\ln(k)$. Then, two types of regimes appear: linear, $\eta/\eta_0 \sim 0.13\phi_r$ when $\phi_r \leq \phi_r^*$, and a power law, $\eta/\eta_0 \sim \phi_r^{0.48}$, for $\phi_r > \phi_r^*$, with $\phi_r^* = 5$. Even for low $\phi_r$ the influence of the bonding interactions are stronger on the viscosity than changing the molecular weight of the copolymer branches or their persistence length, because the slope is over four times larger when changing *k*. Knowing the effect of *k* on the viscosity should be useful when synthesizing new thickeners for various applications, taking advantage of ionic and covalent bonding interactions, for example.

Our results indicate that forming large associations of thickening polymers in complex fluids is a key factor to enhance their viscosity. This conclusion agrees with experiments on some star Telechelic ionomers [25 – 27], showing that their viscoelastic properties can be controlled simply by modifying the number of ionic groups per star ionomer. Doing so favors the formation of networks between ionomers, which can be ascertained by the rheological data. The ionomers couple via dipolar connections to create elastically effective sections of the network. Polymer intermixing and emergence of networks are favored when noncovalent bondings are present. This insight is important in the design and optimization of new, efficient low molecular weight viscosifying polymers. The solvent used in this work is a generic model solvent so that our conclusions can be equally useful to aqueous thickening polymers and also to those used to viscosify nonpolar fluids, such as supercritical $CO_2$.

## ACKNOWLEDGMENTS

We acknowledge funding from the European Union's Horizon 2020 Program under the ENERXICO Project, grant agreement No. 828947 and under CONACYT-SENER-Hidrocarburos (Mexico), grant agreement No. B-S-69926. The calculations reported here were mostly performed using the supercomputing facilities of ABACUS Laboratorio de Matemática Aplicada y Cómputo de Alto Rendimiento of CINVESTAV-IPN. The authors acknowledge also the computer resources offered by the Laboratorio de Supercómputo y Visualización en Paralelo (LSVP − Yoltla) of UAM − Iztapalapa, where part of the simulations were carried out.

48

## TOC GRAPHIC

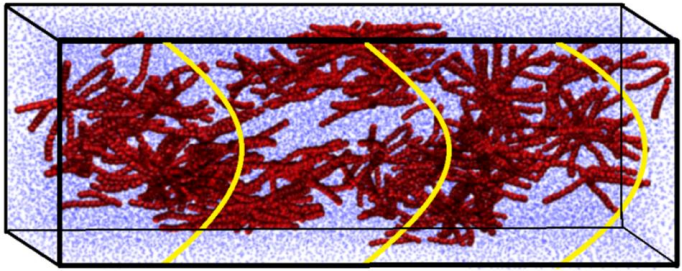
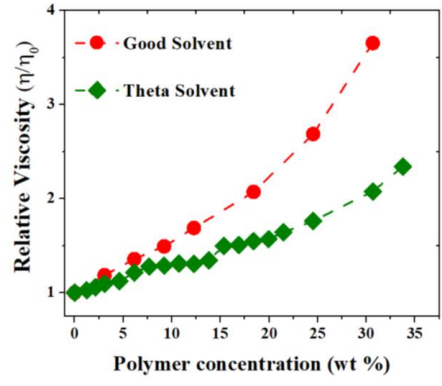